\UseRawInputEncoding
\documentclass[
preprint,
superscriptaddress,
showpacs,preprintnumbers,
 amsmath,amssymb,
 aps,
 prx,
floatfix,
]{revtex4-2}

\usepackage{graphicx}
\usepackage{dcolumn}
\usepackage{bm}

\newcommand{\wsefull}[0]{$2$H-WSe$_2$}
\newcommand{\wse}[0]{WSe$_2$}

\newcommand{\inva}[0]{\AA{}$^{-1}$}

\newcommand{\KV}[0]{$\overline{K}$ valley}
\newcommand{\KP}[0]{$\overline{K}$ point}

\newcommand{\SV}[0]{$\overline{\Sigma}$ valley}
\newcommand{\SP}[0]{$\overline{\Sigma}$ point}
\newcommand{\GP}[0]{$\overline{\Gamma}$ point}

\newcommand{\GK}[0]{$\overline{\Gamma}$-$\overline{K}$}
\newcommand{\flu}[0]{$\mu$J/cm$^{2}$}
\newcommand{\bGSK}[0]{$\overline{\Gamma}$-$\overline{\Sigma}$-$\overline{K}$}
\newcommand{\bKM}[0]{$\overline{K}$-$\overline{M}$}
\newcommand{\bMG}[0]{$\overline{M}$-$\overline{\Gamma}$}

\newcommand{\GW}[0]{G$_0$W$_0$}
\begin{document}


\title{Excited-state band structure mapping}

\author{M. Puppin}%
\affiliation{Laboratoire de Spectroscopie Ultrarapide and Lausanne Centre for Ultrafast Science (LACUS), École Polytechnique Fédérale de Lausanne, ISIC, Station 6, CH-1015 Lausanne, Switzerland}

\affiliation{Fritz-Haber-Institut der Max-Planck-Gesellschaft, Faradayweg 4-6, 14195 Berlin, Germany}%

\email{michele.puppin@epfl.ch}

\author{C. W. Nicholson}%
\author{C. Monney}%
\affiliation{Department of physics and Fribourg centre for nanomaterials, University of Fribourg, Chemin du Musée 3, CH-1700 Switzerand}%

\author{Y. Deng}%
\affiliation{Paul Scherrer Institute, SwissFEL, 5232 Villigen PSI, Switzerland}%

\author{R. P. Xian}%
\author{J. Feldl}%
\author{S. Dong}%
\affiliation{Fritz-Haber-Institut der Max-Planck-Gesellschaft, Faradayweg 4-6, 14195 Berlin, Germany}%

\author{A. Dominguez}%
\affiliation{Shenzhen JL Computational Science and Applied Research Institute, Shenzhen, 518131 China}%
\affiliation{Beijing Computational Science Research Center, Beijing 100193, China}%
\author{H. H{\"u}bener}%
\affiliation{Max Planck Institute for the Structure and Dynamics of Matter and Center for Free Electron Laser Science, Luruper Chaussee 149, Geb. 99 (CFEL), 22761 Hamburg}%
\author{A. Rubio}%
\affiliation{Max Planck Institute for the Structure and Dynamics of Matter and Center for Free Electron Laser Science, Luruper Chaussee 149, Geb. 99 (CFEL), 22761 Hamburg}%
\affiliation{Center for Computational Quantum Physics, Flatiron Institute,
162 5th Ave., New York, 10010 NY, USA}
\affiliation{Nano-Bio Spectroscopy Group, Universidad del Pa{\`i}s Vasco UPV/EHU- 20018 San Sebastián, Spain}

\author{M. Wolf}%
\author{L. Rettig}%
\affiliation{Fritz-Haber-Institut der Max-Planck-Gesellschaft, Faradayweg 4-6, 14195 Berlin, Germany}%
\author{R. Ernstorfer}%
\affiliation{Fritz-Haber-Institut der Max-Planck-Gesellschaft, Faradayweg 4-6, 14195 Berlin, Germany}%
\affiliation{Institut f{\"u}r Optik und Atomare Physik, Technische Universität Berlin, Straße des 17.~Juni 135, 10632 Berlin, Germany}%

\email{ernstorfer@fhi-berlin.mpg.de}

\date{\today}

\begin{abstract}
Angle-resolved photoelectron spectroscopy is an extremely powerful probe of materials to access the occupied electronic structure with energy and momentum resolution. However, it remains blind to those dynamic states above the Fermi level that determine technologically relevant transport properties. In this work, we extend band structure mapping into the unoccupied states and across the entire Brillouin zone by using a state-of-the-art high repetition rate, extreme ultraviolet femtosecond light source to probe optically excited samples. The wide-ranging applicability and power of this approach are demonstrated by measurements on the 2D semiconductor \wse, where the energy-momentum dispersion of valence and conduction bands are observed in a single experiment. This provides a direct momentum-resolved view not only on the complete out-of-equilibrium band gap but also on its renormalization induced by electron-hole interaction and screening. Our work establishes a new benchmark for measuring the band structure of materials, with direct access to the energy-momentum dispersion of the excited-state spectral function.
\end{abstract}

\maketitle


Functionality in electronic and optoelectronic devices is based on the control of the flow of charge carriers under out-of-equilibrium conditions. At the microscopic level, charge transport and device operation rely upon generating non-equilibrium electron distributions controlled by external fields to achieve the desired electronic response. The propagation of electrons in a crystal and the evolution of their energy distributions are governed by the details of the electronic structure as well as the efficiency of elastic and inelastic scattering processes. 

Time-resolved ARPES (trARPES) addresses this problem by observing the spectral function of a material after excitation via a femtosecond optical pulse \cite{Smallwood2016a}.  The momentum-resolved distribution of excited states combined with the dynamical information on state lifetimes provides a powerful view into excited solids \cite{dongmeasurement}, extending the scope of ARPES and allowing to observe out-of-equilibrium electronic properties which can be used to extract the electronic coupling with phonons and other degrees of freedom \cite{umberto2020,Damascelli2019}. Ultimately, understanding matter out-of-equilibrium is mandatory for achieving optical control in complex materials \cite{ahn_designing_2021}.
TrARPES can resolve states unoccupied at equilibrium, and has been used to reveal the unoccupied band structure of topological materials \cite{sobota_direct_2013}, to measure optically-dressed states \cite{wang_observation_2013}, to observe spin-valley polarizations in the conduction band of transition metal dichalcogenide semiconductors \cite{Bertoni2016} and has enabled the direct observation of excitons \cite{madeo_directly_2020,man_experimental_2021,dongmeasurement}. 

An important open question is how band properties extracted from the trARPES spectral function in the excited state compare with conventional steady-state experiments, e.g.~optical spectroscopy or ARPES. A common expectation is that a comparison is possible in the weak excitation limit \cite{buss_setup_2019} where trARPES experiment become very challenging, particularly when accessing the full Brillouin zone (BZ) of the investigated material. This is beyond the reach of most trARPES experiments, which are performed at ultraviolet (UV) photon energies. Extending these experiments to the extreme-ultraviolet (XUV) photon energy range and correspondingly to high photoelectron momenta covering the whole BZ, while retaining a comparable signal-to-noise ratio and weak excitation densities have been challenging until the recent development of suitable high-repetition-rate XUV sources \cite{puppin_time-_2019,buss_setup_2019,sie_time-resolved_2019,cucini_coherent_2020,liu_extreme_2020}. 

In this work, we employ a state-of-the-art experimental setup \cite{puppin_time-_2019} to simultaneously determine the energy of conduction states (unoccupied at equilibrium) and valence states. This allows us to address the band gap, one of the fundamental opto-electronic properties, by mapping in reciprocal space both valence and conduction bands of \wsefull, a two-dimensional transition metal dichalcogenide (TMD) semiconductor widely studied for excitonic and spin-valleytronic applications \cite{ciarrocchi_polarization_2019,unuchek_valley-polarized_2019,wang_electronics_2012}. 

The conduction band population is probed with a 21.7 eV XUV pulse following photoexcitation by a 3.1 eV pulse, with a temporal resolution better than 100 fs. Excited-state ARPES measurements are performed before energy relaxation to the conduction band minimum and reveal the energy versus momentum dispersion of valence and conduction states in a single experiment. 
We then study the excited-state band gap and its renormalization due to many-body effects and demonstrate that in the low-excitation limit the trARPES gap agrees with the band gap measured by other spectroscopies and predicted by theory. This validates excited-state band structure mapping as a generally applicable method to measure, with momentum resolution, the conduction states of materials. 



To better understand the difference and similarities between ARPES and trARPES, we shortly review the two experimental approaches. In an ARPES experiment, a photon with energy $h\nu$ excites a single-crystalline sample, and the kinetic energy $E$ of photoelectrons is measured along a wavevector direction $\textbf{k}$. If photoionization is treated as a sudden process, the photoemission intensity can be approximated as \cite{Damascelli2003}: 

\begin{equation}\label{eq:SpF}
I(\textbf{k},E) = I_0(\textbf{k},E) A^{-}(\textbf{k},E) f_{\mu,T}(E).
\end{equation}

\begin{figure}[ht]
\includegraphics[width=1.00\columnwidth]{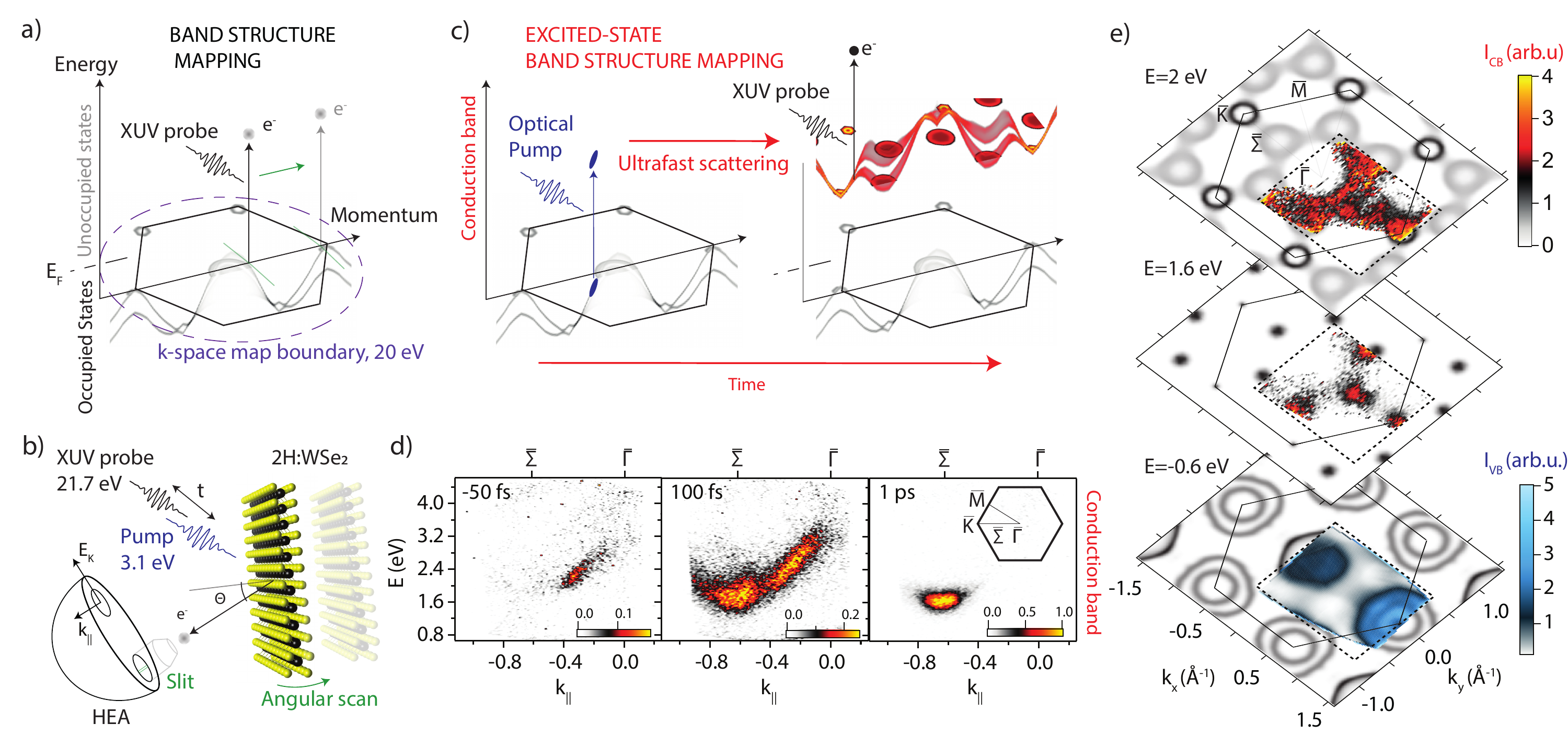}
\caption{\label{fig:figure1} a) Band structure mapping in reciprocal space by angle-resolved photoelectron spectroscopy (ARPES). The reciprocal space region measured by the hemispherical energy analyser (HEA) for two sample tilt angles is indicated by a green line, the maximum parallel momentum which can be accessed by 20 eV photons is indicated by a violet dashed line. b) trARPES experiments on \wsefull: an optical pump pulse at an energy of 3.1 eV excites the system. At a delay t, an XUV probe pulse at an energy of 21.7 eV generates photoelectrons, which are measured as a function of the emission angle $\theta$ with a HEA. The sample angle is scanned across the analyser slit to collect ARPES maps. c) Excited-state band structure mapping d) trARPES data collected in the conduction band of \wse\ for pump-probe delays of -50 fs, 100 fs and 1 ps. Inset: the surface Brillouin zone of \wse. e) Photoelectron intensity distribution as a function of parallel momentum for three energies at a pump-probe delay of 100 fs; VB and CB energy distribution curves have been independently intensity normalized for better visualization. The experimental data is collected in a region delimited by the dashed line. Outside this region, the results of \GW\ calculations are displayed, the theoretical bands dispersion along the $k_z$ direction was integrated; the conduction bands was rigidly offset by a \textit{scissor operator} to match the experimental energy.}
\end{figure}

Equation \ref{eq:SpF}, which for simplicity neglects the experimentally finite angular and energy resolution, as well as charge transport at the surface, links the ARPES spectrum $I(\textbf{k},E)$ to the underlying electronic structure via three factors. The one-electron-removal spectral function, $A^{-}(\textbf{k},E)$, contains the information about the quasi-particle band structure and many-body interactions. The spectral weight is modulated by a matrix element term $I_0(\textbf{k},E)$, which depends on initial and final state symmetry and wave vectors, as well as photon energy ($h \nu$) and polarization, and the experimental geometry \cite{Moser2016,beaulieu_revealing_2020}. Thirdly, the Fermi-Dirac distribution $ f_{\mu,T}(E) $ imposes that only states populated at the temperature T can contribute to the measured spectrum, setting a limit to the highest accessible energy to few $ k_B T $ above the chemical potential $\mu$. The matrix element term is vanishing unless momentum conservation parallel to the sample's surface is fulfilled by the escaping photoelectron, allowing to link the measured photoelectron angular distribution $I(\textbf{k},E)$ to the quasi-particle bands in reciprocal space, as illustrated in Fig. \ref{fig:figure1} a). Parallel momentum ($\textbf{k}_{\parallel}$) conservation, together with energy conservation, imposes that typically only energetic photons in the XUV range can access the whole BZ \cite{hufner2003photoelectron}. As an example, photons with an energy of $\approx 20$ eV are necessary to measure the first BZ boundary of \wse, as indicated by the violet dashed line in Fig. \ref{fig:figure1} a). In our experiment photoelectron spectra are collected with a hemispherical energy analyser (HEA) which measures kinetic energy ($E_K$) and angle of emission along the entrance slit (Fig. \ref{fig:figure1} b), this corresponds to a line-cut throughout the function $I(\textbf{k},E)$ (full green lines in Fig. \ref{fig:figure1} a). Band mapping is achieved by angular scanning of the sample (green arrows in Fig. \ref{fig:figure1} a) and b) across the analyser slit. The multidimensional function $I(\textbf{k},E)$ is constructed from different images and data can be displayed as constant energy cuts or as energy versus momentum plots, as shown in Fig. \ref{fig:figure1} a) where a horizontal constant energy cut close to the valence band maximum and a vertical energy versus momentum dispersion across the BZ are plotted. It is worth noting the alternative approach of momentum microscopy, in which the whole accessible photoemission space is collected at the same time \cite{kotsugi_microspectroscopic_2003}. A detailed comparison between the two methods reveals that an HEA ensures higher counting statistics when acquiring data along a specific direction \cite{maklar_quantitative_2020}, whereas the fixed geometry provided by momentum microscopy is suitable for the study of the symmetry-dependent matrix element $I_0(\textbf{k},E)$ \cite{beaulieu_revealing_2020}.  

A time-resolved ARPES experiment accesses an excited state of the material by performing an ARPES experiment at a well-defined temporal delay t following a femtosecond optical pump pulse (\ref{fig:figure1} b). The trARPES spectrum $\tilde{I}(\textbf{k},E,t)$ thereby measures the (quasi)-electron-removal spectrum as a function of this time delay:

\begin{equation}\label{eq:SpF_TR}
\tilde{I}(\textbf{k},E,t) = \tilde{I_0}(\textbf{k},E,t) \tilde{A}^{-}(\textbf{k},E,t) \tilde{f}(\textbf{k},E,t)   .
\end{equation}

Here eq. \ref{eq:SpF} is modified to include the explicit time dependence of each term. The optical excitation produces not only an out-of-equilibrium electronic distribution $\tilde{f}$, but also perturbs the many-body interactions in the spectral term $\tilde{A}^{-}$. The matrix element term $\tilde{I_0}$ can become a time-dependent quantity if the symmetry of the initial or final states is modified \cite{boschini_role_2020}. We follow the convention that for $t>0$ the pump excitation occurs before photoemission: recovery of equilibrium requires that  $\tilde{I}(\textbf{k},E,t) \xrightarrow{t \rightarrow +\infty} I(\textbf{k},E)$. 

As illustrated in Fig. \ref{fig:figure1} b, trARPES provide access to states unoccupied at equilibrium. This can be understood as a two-step process, where, in a first step the femtosecond pump pulse creates an optical polarization in allowed momentum and energy regions, corresponding to vertical optical transitions in the material ($\Delta \mathbf{k}=0$) \cite{koch_semiconductor_2006}. In a second step, microscopic scattering events within a few hundred femtoseconds redistribute electronic population to multiple states across the conduction band (CB) (Fig. \ref{fig:figure1} a). 
Electrons relax their excess energy via multiple electron-phonon scattering events towards the band edges and accumulate at the CB minima on time scales typically shorter than a few picoseconds. By measuring the photoelectron energy and angular distribution before significant energy relaxation to the lattice has occurred, the information encoded in $\tilde{A}^{-}$ can be revealed in a range $E<\mu + h \nu_p$, where $h \nu_p$ is the pump photon energy. 

Excited-state band mapping of unoccupied states is particularly demanding and strongly benefits from high repetition rate ($>$ 100 kHz) XUV sources. First, a sufficiently short XUV pulse is fundamental for accessing the out-of-equilibrium state before its decay throughout the BZ. In addition, space charge effects, which are inherent in ARPES with short XUV pulses, are mitigated in high repetition rate experiments \cite{hellmann_vacuum_2009}. Furthermore, the higher the pump excitation energy density, the stronger many-body interactions modify the function $\tilde{I}(\textbf{k},E,t)$ relative to the equilibrium case. trARPES experiments at high-repetition rates benefit from higher counting statistics and hence data can be acquired at weaker perturbation strength. 

To meet the simultaneous requirements of an ultrashort XUV source with a high repetition rate, in this work we generate probe pulses by high-harmonic generation with an optical parametric chirped pulse amplifier operating at 500 kHz \cite{Puppin2015}. This results in XUV pulses at an energy of 21.7~eV and with characteristic time-bandwidth product of approximately 20~fs$\times$110~meV \cite{puppin_time-_2019}, which are temporally short enough to access the excited states before significant carrier energy relaxation has occurred and, at the same time, have an energy bandwidth sufficiently narrow to resolve the excited-state energy features. trARPES experiments were performed on single-crystalline samples of bulk \wse, cleaved in ultra-high vacuum conditions. The material was excited by a pump pulse with a photon energy of 3.1~eV and at an excitation energy density of 40~\flu. 

To illustrate the ability of trARPES to visualize states which are unoccupied at equilibrium, we show in Fig.~\ref{fig:figure1} d) energy versus momentum data collected in an energy window in the conduction band (CB) along the high symmetry direction \GK. Three selected time delays (-50~fs, 100~fs and 1~ps) are plotted side by side. The surface BZ of \wse, with the high symmetry points marked, is shown as an inset of Fig.~\ref{fig:figure1} d). During the rising edge of the pump pulse (-50~fs), the CB signal is localized at -0.35~\inva\ from the BZ center (\GP). This suggests that in this region population is transferred via an optical transition at the photon energy of 3.1~eV, rather than indirectly by scattering. The intensity of this feature as a function of time was used as a measure of the pump-probe temporal cross-correlation and the temporal maximum was used to define the time zero. The full-width at half maximum of the cross-correlation is 95~fs, dominated by the pump pulse duration \cite{note:SI}. Throughout this work, the zero energy was set for convenience to the valence band energy at the \KP, the corner of the hexagonal BZ. 

At a time delay of 100~fs, population can be observed throughout the conduction states, up to at an energy $\approx$ 2.5~eV (Fig. \ref{fig:figure1} d), central panel). This delay was selected to perform the excited-state band structure mapping. Relaxation towards the $\Sigma$ conduction band valley minimum is indeed already apparent at a delay of 1 ps (Fig. \ref{fig:figure1} d), right panel). 

An energy window from -1.5 to 3.5~eV was selected to observe simultaneously valence and conduction bands around the band gap, which is a unique feature of trARPES. Three exemplary constant energy cuts of the data at $t=100$ fs are shown in Fig. \ref{fig:figure1} e), which display in false colours the photoelectron intensity distribution as a function of parallel momentum for energies of -0.6 eV in the valence band (VB), 1.6 eV and 2 eV in the conduction band (CB). The measurement region is indicated by a dashed line and comprises the whole first BZ of \wse. Two different false color scales are used for conduction and valence states; energy distribution curves were normalized independently in the CB and VB for a clearer display of the constant energy maps \cite{note:SI}.

To rationalize the experimental data we perform \textit{ab initio} density functional theory (DFT) calculations of the electronic band structure using the generalised gradient approximation with the PBE functional, as implemented in the QUANTUM ESPRESSO package \cite{Giannozzi2009}. To improve the agreement with experimental data, we use many-body perturbation theory at the one-shot \GW\ level \cite{hedin_new_1965,GWmethod} on top of DFT results \cite{note:SI}. This computes quasiparticle energies, correcting to lowest order the unscreened electronic Green's function $G_0$ by the Coulomb interaction $W_0$. The quasi-particle energy dispersion is calculated as a function of the three-dimensional wave-vector ($k_x$,$k_y$,$k_z$). For a direct comparison with data in Fig. \ref{fig:figure1} d), the theoretical bands are integrated along the reciprocal space direction orthogonal to the sample surface ($k_z$). This choice is justified by the strong surface sensitivity of XUV-based photoemission due to the short mean-free-path of photoelectrons. Electron momentum conservation is relaxed for the $k_z$ component, adding an additional source of energy broadening for bands with dispersion out of the surface plane. There is strong evidence that in \wse\ the photoemission probing depth at 21.7 eV is mostly limited to the uppermost layer ($\approx$ 0.5 nm), in fact, inversion-symmetric \wse\ surprisingly exhibits strong spin-polarized bands \cite{Riley2014c} and valley polarization in circularly-pumped tr-ARPES \cite{Bertoni2016}. The importance of final state effects in the material is evidenced by one-step photoemission calculations \cite{beaulieu_revealing_2020}, and will be discussed further below.

\begin{figure}[ht]
\includegraphics[width=1.0\columnwidth]{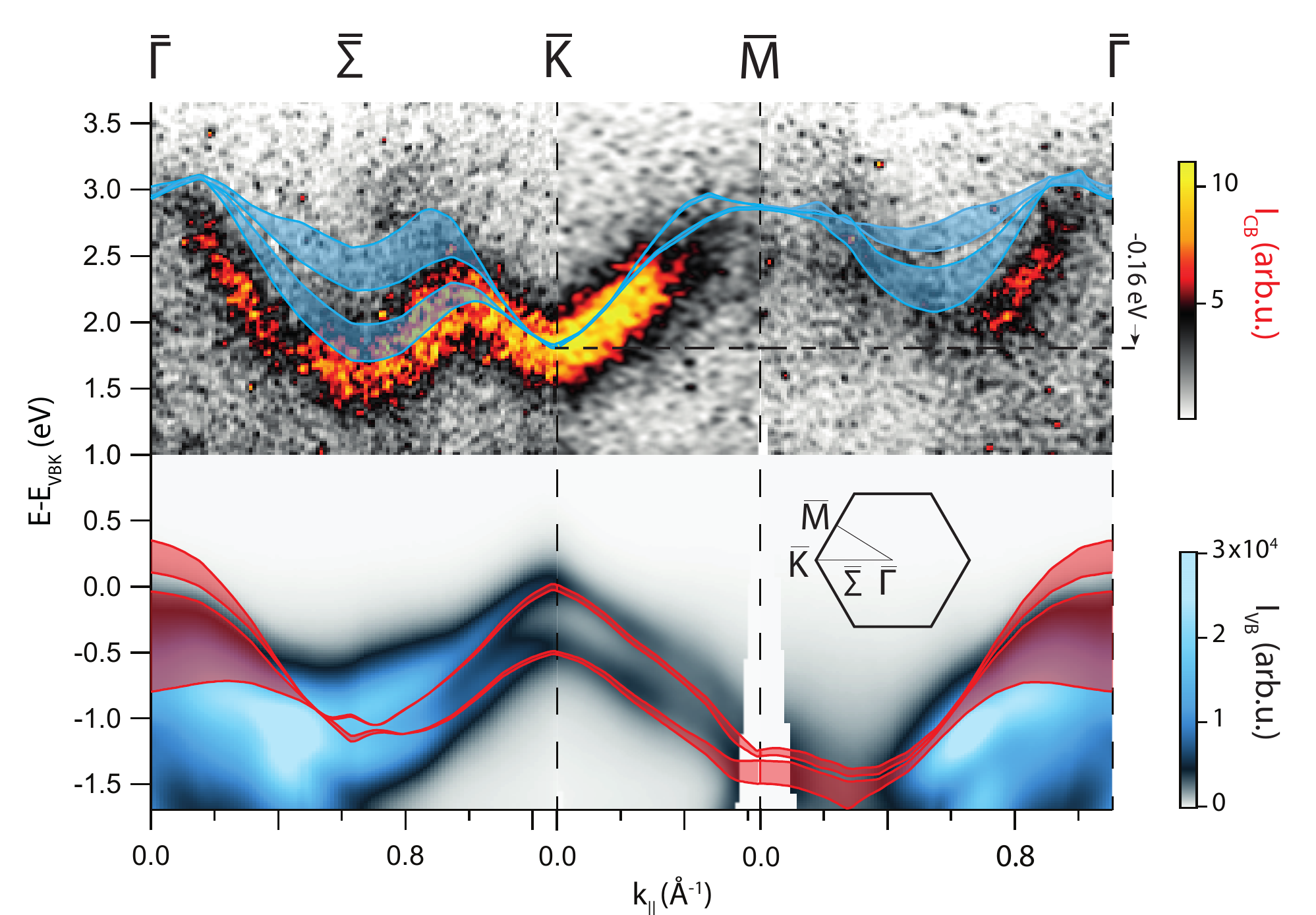}
\caption{\label{fig:figure3} Measured ARPES intensity as a function of energy and parallel momentum showing the VB and CB along the \bGSK, \bKM, \bMG\ directions, indicated in the upper panel. Conduction band states are displayed by a different color scale. Blue and red curves indicate the quasiparticle energies calculated with the \GW\ method for the CB and VB, respectively. The theoretical band structure energy zero was set to the VB position at the \KP, the CBs (blue) were rigidly shifted by a -0.16 eV \textit{scissor operator} to match the \SV\ center energy. The momentum dispersion along the $k_z$ direction is indicated by the shaded area.}
\end{figure}

The experimental data contains the excited-state CB and VB energy-momentum dispersion for arbitrary reciprocal space directions, which can be compared with our \textit{ab initio} calculations and with other experiments. For this purpose, energy versus momentum photoelectron distributions are plotted along three high-symmetry directions \bGSK, \bKM, \bMG\ in Fig. \ref{fig:figure3} and compared with the results of the calculations. The theoretical $k_z$ dispersion is indicated by a shading, highlighting two-dimensional (low $k_z$ dispersion) and three-dimensional states. The experimental photoelectron intensity is plotted without additional normalization, and intensity modulations are attributable to the momentum dependent matrix element. The average intensity of the conduction band signal is a factor 10$^{-3}$ that of the valence states, and we use two distinct false color scales for conduction and valence states, respectively. 

The zero energy reference is set to the highest energy VB at the $\overline{K}$ point also for the theoretical data, to minimize any alignment uncertainty due to $k_z$ dispersion. The theoretical conduction states were rigidly shifted by -160 meV to match the measured CB energy at the \KP, both in Fig. \ref{fig:figure1} d) and in Fig. \ref{fig:figure3}.

Theory predicts two valence and two conduction bands in the observed energy window, as all calculated bands are spin-degenerate, consistent with the inversion-symmetric bulk crystal structure of \wsefull. The spin-orbit splitting of the VB band at the \KP\ is $\approx$ 500 meV, in good agreement with past literature \cite{Finteis1997a,Riley2015a,Tanabe2016}. Despite being a layered quasi-2D material, \wse\ displays some inherently three-dimensional features. In particular, the $\Sigma$ valley, as well as the valence band at the $\overline{\Gamma}$ point, have considerable $k_z$ dispersion. In contrast, the out-of-plane band dispersion is low in the vicinity of the $\overline{K}$ point, as confirmed by energetically-narrower features in ARPES. Our \GW\ calculations predict an orthogonal momentum dispersion on the order of 40 meV for the VB and 30 meV for the CB at the $\overline{K}$ point. Calculations place the indirect band gap between the maximum of the VB at the $\overline{\Gamma}$ point and the $\Sigma$ valley. In our data the conduction band minimum (CBM) is unambiguously located  at the $\overline{\Sigma}$ point, however the apparent valence band maximum (VBM) is observed at the $\overline{K}$ point, and a broad continuum of states is observed at the \GP. It is widely accepted that the absolute VB maximum is located at the $\overline{\Gamma}$ point and that matrix element effects cancel the contribution of the upper VB at $\overline{\Gamma}$ \cite{Finteis1997a,Riley2014c}. After the rigid offset of -160 meV mentioned above, the \GW\ calculations are in qualitative agreement with the excited-state band structure and reproduce the main features of the experimental conduction band. 

For a quantitative comparison, the quasi-particle energy must be determined from the ARPES intensity. Final state effects usually complicate the retrieval of quasi-particle energies and of many-body effects in the spectral function. However, the problem is absent in a strictly two-dimensional state (dispersion only along $\textbf{k}_{\parallel}=(k_x,k_y)$) \cite{hufner_very_2007}. Both valence and conduction states at the direct optical band gap at the \KP\ are quasi-two-dimensional, enabling for robust comparison of the experimental excited-state band gap with theory and other experimental techniques.

\begin{figure}[ht]
\includegraphics[width=1\columnwidth]{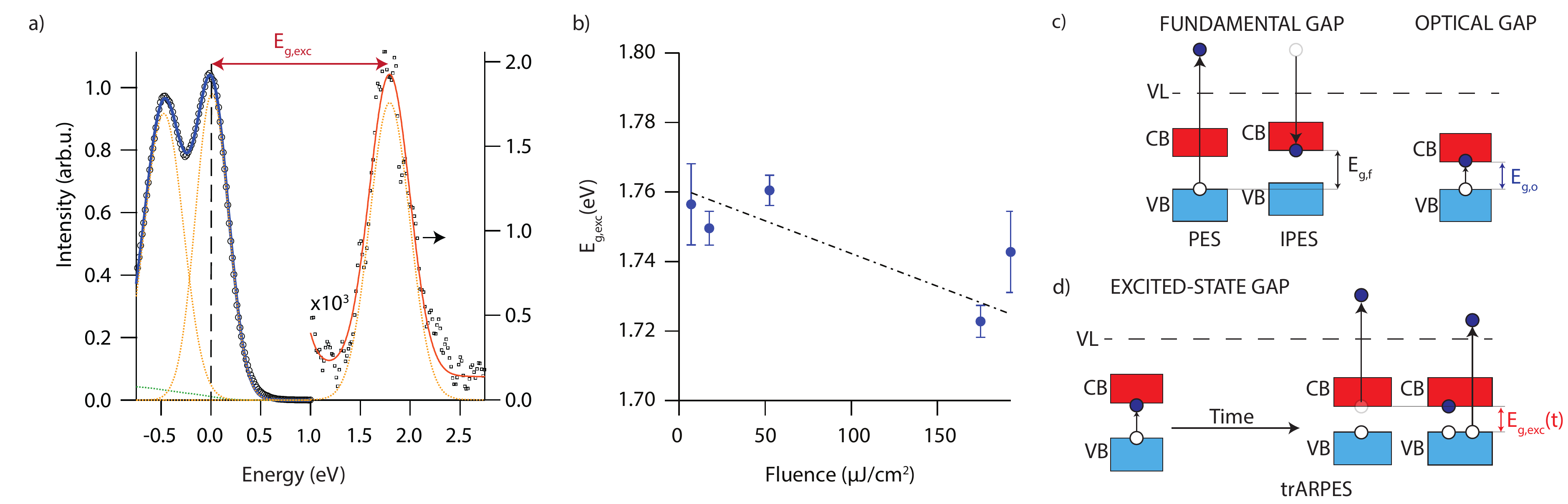}
\caption{\label{fig:figure4} a) Energy distribution curve at the K point, together with the fit used to determine the excited-state band gap $E_{g,exc}$. The conduction band signal intensity, displayed on the right-hand axis, was scaled by a factor $10^3$ for clarity. b)  Fluence dependence of the excited-state band gap. c) Schematic comparison between fundamental, optical, and excited-state band gaps, VL indicates the vacuum level. }
\end{figure}

The CB and VB energies are extracted from the experimental data by a fit of the energy distribution curve (EDC) at the \KP, for $t=100$ fs. The procedure is illustrated in Fig. \ref{fig:figure4} a), the photoelectron spectrum of the VB is well fitted by two Gaussian peaks, and by a Shirley background. The two, nearly-degenerate conduction bands predicted by theory are not resolved within the experimental line width, and a single Gaussian peak describes well the CB signal. Due to its higher intensity, the higher energy tail of the VB spectrum appears as a background on the CB, and is modeled by an exponential decay. We define the experimental band gap as the distance between the uppermost VB peak position (E=0 by definition) to the center of the CB peak, as highlighted by the red line in Fig. \ref{fig:figure4} a) and we measure a band gap of 1.76$\pm$0.01 eV. We note that this procedure, valid for quasi-2D bands, differs from the method adopted for three-dimensional semiconductors, where the band edge is found by linear extrapolation of the photoelectron spectral edge \cite{katnani_microscopic_1983}. 

The excited-state quasi-particle energy, an out-of-equilibrium quantity, can change as a function of the excitation energy density \cite{roth_photocarrier-induced_2019}. To investigate the impact on the band gap, we follow its evolution for increasing incident optical energy density up to 200 \flu\ and observe a decrease of the band gap (Fig. \ref{fig:figure4} b). The maximum effect is $\approx$ 50 meV, with a linear slope of $1.8\times10^{-1}$ meV/($\mu$J/cm$^2$); the extrapolated limit at zero excitation density is 1.76$\pm$0.03 eV. 

It is interesting to compare this experimental band gap, which we call the \textit{excited-state band gap} $E_{g,exc}$, with ab-initio calculations and other experimental techniques. Several experiments have been designed to resolve the electronic structure above the chemical potential \cite{fuggle1992unoccupied}. Inverse photoemission \cite{himpsel_inverse_1990}, scanning tunneling spectroscopy \cite{STSref}, and very low-energy electron diffraction \cite{strocov_very_2000} access unoccupied conduction states by adding an electron to the system and probing the complementary one-electron-addition spectral function $A^{+}(\textbf{k},E)$ \cite{seitz_effects_1970}. Angle-resolved inverse photoemission (ARIPES), in particular, has momentum resolution \cite{fuggle1992unoccupied}. Unfortunately, due to the small cross-section of the process and, unlike ARPES, due to the lack of parallel detectors with multiple angular and energy channels, ARIPES has not evolved to a similarly widespread technique \cite{himpsel_inverse_1990}. Another approach can used in photoemission to observe otherwise unoccupied states, namely sample doping by alkali metal atoms \cite{Riley2015a,Kim2016_corrected}. A limitation of alkali doping is the possibility of chemical modification to the band structure \cite{Riley2015a}. Additionally, resonant inelastic X-ray scattering techniques have also been used to map the dispersion of unoccupied states \cite{monney_mapping_2020,monney_mapping_2012}.  The direct gap at the \KP\ for \wse\ from various methods is displayed in table \ref{table:1}.


\begin{table}[h!]
\centering
\begin{tabular}{|c| c | c| c|} 
\hline
\textbf{Method} & \textbf{Band gap (eV)} & Reference\\  [0.5ex] 
\hline
ARPES+ARIPES & 1.7, 1.4  & \cite{Finteis1997a},\cite{Traving1997} \\  
ARPES+Doping & 1.62  &\cite{Kim2016_corrected} \\  
trARPES & \textbf{1.76}  & This work \\  
Optics, A-exciton & 1.697$^{*}$, 1.60, 1.626 &  \cite{Beal1976e}, \cite{zeng_optical_2013}, \cite{Arora2015} \\ 
Optics, Interband & 1.752$^{*}$, 1.686 & \cite{Beal1976e}, \cite{Arora2015}  \\ [1ex] EELS, A-exciton & 1.75  & \cite{Schuster2016}  \\ [1ex]
\hline\hline
DFT & 1.25, 1.17-1.55 & This work, \cite{jain_materials_2013,Roldan2014,kumar_electronic_2012,huang_theoretical_2014,curtarolo_aflowliborg_2012} \\ 
 $G_0W_0$ & \textbf{1.90}, 1.75, 2.08$^{\ddag}$ &  This work, \cite{Jiang2012}, \cite{he2014a}\\  
 BSE, A-exciton & 1.86$^{\ddag}$ &  \cite{he2014a} \\  
 BSE, Interband & 2.02$^{\ddag}$ &  \cite{he2014a} \\  
 \hline
\end{tabular}
\caption{Comparison between experimental (upper part) and theoretical band gap of \wse\ (lower part) at the \KP\ (direct band gap) . $^{*}$Measured at 77 K, at room temperature the gap is reduced by $\approx$60 meV \cite{Arora2015} . $^{\ddag}$ bilayer \wse.} 
\label{table:1}
\end{table}

The fundamental or quasiparticle band gap $E_{g,f}$ is usually defined as the difference between the electron affinity, i.e.~the energy gained by adding a single electron to an N electron system, and the ionization energy, needed to remove an electron leaving N-1 electrons behind \cite{jbaerends_kohnsham_2013}. The quasiparticle gap should not be confused with the so-called optical band gap, which will be discussed later on. The so-called transport band gap, determined by electrical transport measurements, coincides with the fundamental band gap, however, in the case of semiconductors such as bulk \wse, possessing an indirect band gap and multiple conduction band valleys, momentum-resolved techniques provide a more complete picture. In view of comparison with optical spectroscopy, we restrict here to the case of the direct band gap and we more loosely consider the band gap as a momentum-dependent quantity, which attains its minimum at the direct fundamental band gap. 
Experimentally, the momentum-dependent quasiparticle band gap can be measured by comparing the VB measured by photoemission (N-1 electron final state) with the CB measured by inverse photoemission (N+1 electrons final state). This procedure is schematized in Fig. \ref{fig:figure4} c) and necessitates a common energy reference between the two experimental setups. In particular, the direct fundamental gap of \wse\ at the \KP\ was experimentally measured to be $E_{g,f}^{exp}=1.7\pm0.1$ eV by combining ARPES and ARIPES \cite{Finteis1997a}.

When comparing the experimental gap with theoretical results, an important question is to what extent one is allowed to compare \textit{ab initio} calculations such as DFT with energies determined by (time-resolved) photoelectron spectroscopy. 
DFT computes the ground state electronic density and returns a set of self-consistent Kohn-Sham (KS) bands \cite{giustino2014materials}. 
Even in an idealized case where the exact density functional is known, a direct comparison between the KS bands and the ARPES measurements is not justified \cite{perdew_physical_1983}. Nonetheless in many cases, within a constant energy offset, the KS bands are in good agreement with ARPES data of the valence band. For \wse, in particular, DFT bands reproduce reasonably well the ARPES VB energy dispersion \cite{Riley2014c,Finteis1997a,Tanabe2016,Straub1996}. 
However, if $E_{g,f}$ is directly calculated from the KS bands, theory grossly underestimates the band gap. Before applying the \GW\ correction, our calculations predict a gap value of 1.25~eV, in line with other DFT results, reported in table \ref{table:1}. 
This well-known \textit{band-gap problem} is intrinsic to DFT \cite{Perdew2009}, and is a reminder that KS energies are indeed not quasi-particle energies. Conversely, Hedin's GW method \cite{GWmethod,Hybertsen1986} can be used to calculate quasiparticle excitations in a solid, such as measured in ARPES (electron removal) or ARIPES (electron addition). 
GW calculations correct the DFT energies by an approximate electronic self-energy, typically performed to the lowest order (\GW). We find a considerable improvement in the calculated fundamental gap and obtain a value $E_{g,f}^{GW}= 1.90$ eV, in line with previous calculations \cite{Jiang2012}. 

A second commonly defined band gap is the so-called \textit{optical band gap} $E_{g,o}$, which corresponds to the lowest energy required for a vertical ($\Delta\textbf{k}=0$) electronic transition in the system (Fig. \ref{fig:figure4} c). This is a neutral excitation where both the initial and final states have N electrons, in contrast with the case of the fundamental gap, which is calculated as the energy difference between an N+1 and an N-1 electrons state. The optical band gap is experimentally measured by optical absorption spectroscopy. A remarkable feature in optical absorption spectra is the appearance of excitonic resonances at energies below the onset of electronic interband transitions. The observation of an excitonic peak is the hallmark of the electron-hole interaction, and its center energy defines the optical band gap. To predict the optical band gap one must solve the Bethe-Salpeter equation \cite{rohlfing_electron-hole_2000}. In the optical absorption spectra of bulk \wse\ the so-called A exciton is the lowest resonance at an energy of 1.68 eV, the exciton binding energy $E_x$ was determined to be 50 meV, and the inter-band transition has an energy 1.73 eV \cite{Beal1976e}. This sets the scale for the electron-hole interaction in bulk TMD semiconductors, and one expects $E_{g,o} \approx E_{g,f}-E_x$.

In the \textit{excited-state band gap} measurement (Fig. \ref{fig:figure4} d), a neutral optical excitation is followed by an ionization step at time t, leading to a N-1 electron excited final state with an additional hole in the VB, which is generated for t=0 and is followed by a relaxation dynamics for $t > 0$. The band gap is measured by comparing the kinetic energy of photoelectrons originating from the CB and the VB. Generally speaking, $E_{g,exc}(t)$ is a time-dependent quantity influenced by many-body effects, and can be renormalized by electron-electron interactions, leading to screening and excitonic effects, and by the electron-phonon coupling with the (non-thermal) phonon distribution. 

Our data shows that in the low excitation limit, $E_{g,exc}$(100 fs) is in good numerical agreement with the fundamental band gap determined by other experiments. Furthermore, we observe no signatures of the A excitonic peak at the \KP, which appears in optical measurements at a lower energy of $\approx$ 1.62 eV \cite{Beal1976e,zeng_optical_2013,Arora2015}. A deviation from the single-quasiparticle picture is expected when electron and hole are bound to form excitons \cite{perfetto_first-principles_2016,rustagi_photoemission_2018,christiansen_theory_2019} and photoelectron spectra bear the signature of such interactions as a renormalized energy and momentum dispersion \cite{Weinelt2002,dongmeasurement}. The agreement with the theoretical \GW\ bands in the present case can be rationalized by the fact that the pump photon energy is well above the gap and sufficiently off-resonance to approximate the initial (t $\approx$ 0) carrier distribution as an electron-hole plasma, where exciton quasi-particles are not formed \cite{koch_semiconductor_2006}. In bulk \wse\ the formation of stable A excitons at the \KP\ is hindered by the possibility of electron (hole) scattering to the \SP\ (\GP), which are the global band energy edges. However, if instead the excitation energy is resonant with the excitonic peak observed by optics, excitonic effects can be observed \cite{dongmeasurement}.

We note that \GW\ calculations overestimate the band gap observed in our out-of-equilibrium experiment by $\approx$ 160 meV. However, the agreement with the observed band dispersion is still satisfactory upon a rigid shift of the conduction bands to lower photon energies, suggesting that a single-quasiparticle picture holds well for the excited-state band structure in first approximation. Band gap renormalization is expected to occur due to carrier screening and via electron-phonon coupling \cite{roth_photocarrier-induced_2019,shah2010ultrafast,ulstrup_ultrafast_2016}. Time-resolved diffraction studies reveal that a non-equilibrium phonon distribution rises on the time scale of a few picoseconds \cite{waldecker_momentum-resolved_2017}. At a pump-probe delay of 100 fs, where our data was collected, a significant hot phonon population has not yet developed and we conclude that electronic screening must dominate in band structure mapping experiments and we attribute to this effect the observed band gap reduction at higher excitation densities (Fig. \ref{fig:figure4} b)).

\begin{figure}[ht]
\includegraphics[width=1\columnwidth]{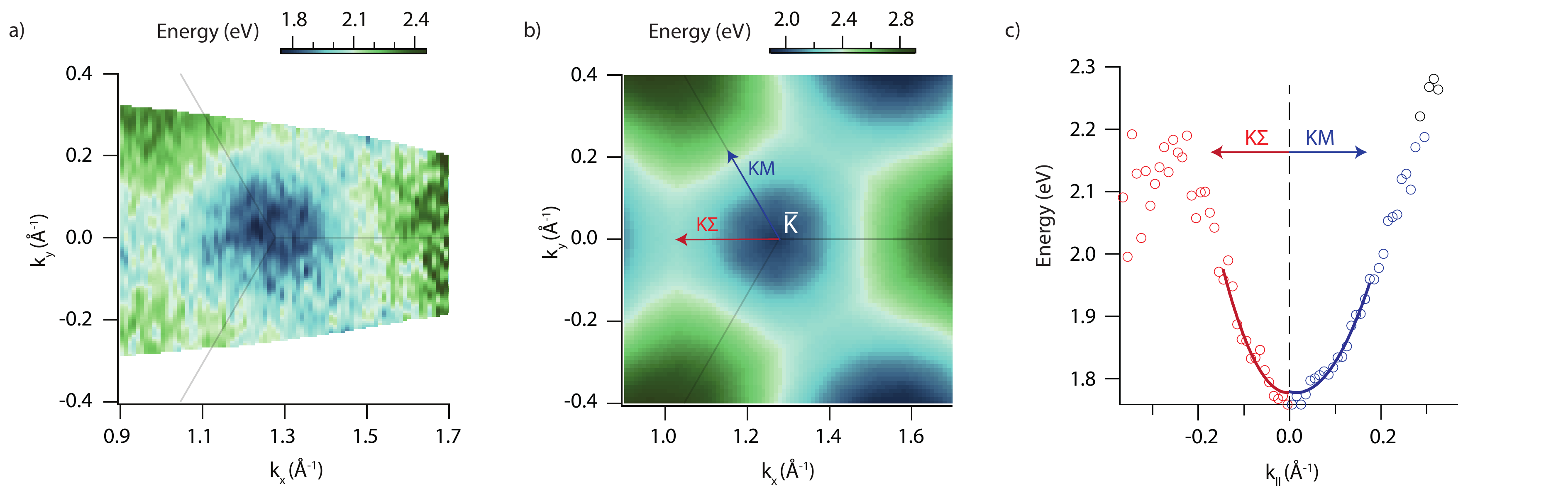}
\caption{\label{fig:figure5}  a) Conduction band center energy at the \KV, b) \GW\ energy of the K valley c) Dispersion along the directions K-$\Sigma$ (negative x-axis) and K-M (positive x-axis). The full line indicates the result of parabolic fits to the data.}
\end{figure}

Having established that the excited-state band gap well approximates the fundamental band gap in our experimental conditions, we now extract the momentum-resolved energy dispersion contained in the experimental maps for the whole 2D \KV. The \KV\ energy is shown in Fig. \ref{fig:figure5} a) and for comparison we plot the theoretical dispersion of the lowest CB in Fig. \ref{fig:figure5} b). The three-fold symmetry of the valley is evident from the data and the anisotropy of the \KV\ can be quantified by extracting the dispersion along the high-symmetry directions K-$\Sigma$ and K-M, indicated in Fig. \ref{fig:figure5} b). For this purpose, we employ the previously described fitting procedure to EDCs surrounding the \KV. The band dispersion of both conduction and valence bands was estimated by fitting a parabola in a range of 0.15 \AA$^{-1}$, as illustrated in Fig. \ref{fig:figure5} c) for the case of the CB. We obtain a value of $m_{e}^{K\Sigma}=0.38\, m_0$ ($m_{h}^{K\Sigma}=-0.52\, m_0$ ) and $m_{e}^{KM}=0.55\, m_0$ ($m_{h}^{KM}=-0.56 \, m_0$ ) for the CB (VB) in the directions K-$\Sigma$ and K-M, respectively, where $m_0$ is the electron mass. The experimental dispersion is somewhat smaller than effective masses reported for DFT, $m_h=-0.625 m_0$ and $m_e=0.821 m_0$ \cite{yun_thickness_2012}. Calculated effective masses from DFT depend strongly on computational details and also on the computational band gap \cite{PRLPOLA}, larger theoretical masses might be therefore linked to the underestimation of the gap in the aforementioned work. 

By observing hole and electron quasi-particle independently, one can calculate effective ($M=m_e+m_h$) and reduced ($\mu_r=m_e m_h/(m_e +m_h )$) exciton masses .
The exciton effective masses are $M^{K\Sigma}=0.9\, m_0$ and $M^{KM}=1.1\, m_0$, which can be compared with experimental results from electron energy loss spectroscopy, $M=0.91\, m_0$ \cite{Schuster2016} and with optical measurements under magnetic field, which report $M=0.7 m_0$ \cite{mitioglu_optical_2015}. The exciton reduced mass determined from our data is $\mu_r^{K\Sigma}=0.22\, m_0$ and $\mu_r^{KM}=0.28\, m_0$. This can be compared with optical absorption spectroscopy data, from which  $\mu_r=0.21\, m_0$ was determined \cite{Beal1976e}. We stress however that, despite the reasonable numerical agreement, other techniques do not identify the hole and electron masses independently. Furthermore, band anisotropy along different symmetry directions can be readily identified and accounted for within the excited-stated band structure. This is particularly relevant for example in valleytronic applications in hetero-layers where energy-degenerate valleys appear at different momentum locations \cite{schaibley_valleytronics_2016}. The detailed effects of layer stacking on the momentum dispersion and on the optical and transport properties is as yet poorly understood and can be directly characterized by excited-state band structure mapping. 

The possibility of visualizing the excited-state band structure by trARPES is demonstrated for the TMD \wse. The experiment provides simultaneous access to valence and conduction states throughout the BZ thereby completely mapping the material's band gap. The excited-state direct gap at the \KP\ agrees in the low-excitation limit with fundamental quasi-particle gap, as obtained by static experiments. Our experiment shows that the excited-state band structure agrees in the low excitation limit with the single-quasiparticle bands and we obtain experimentally conduction and valence band dispersion for the \KP\ for various high symmetry directions. Thanks to XUV light sources at high repetition rate , we anticipate that the measurement of the excited-state band structure in the whole BZ can be performed for a broad class of samples. \GW\ calculations provide a good qualitative description of the data but predict the experimental out-of-equilibrium band gap only within 160 meV. Excite-state band structure mapping can provide an experimental benchmark to quantitatively fine tune computations, e.g. to accurately predict the band gap in high-throughput computational material discovery for optoelectronic applications \cite{rasmussen_computational_2015,curtarolo_aflowliborg_2012}. Automated methods for comparison with theory, demonstrated for multi-dimensional ARPES data \cite{xian2020machine}, are applicable also to excited-state band structure data. Importantly, the method could also provide access to unoccupied states of quantum materials, to resolve topological features above the Fermi level \cite{sobota_direct_2013}, and for correlated materials, e.g. to access the spectral function of unoccupied states in strongly correlated oxides and charge density wave materials \cite{maklar2021nonequilibrium,nicholson_excited-state_2019,nicholson_beyond_2018}. 

\begin{acknowledgments}
This work was funded by the Max-Planck-Gesellschaft, by the German Research Foundation (DFG) within the Emmy Noether program (Grant No. RE 3977/1), and grants FOR1700 (project E5), SPP2244 (project 443366970) and from the European Research Council, Grant Numbers ERC-2015-CoG-682843. M.P. acknowledge financial support by the Swiss National Science Foundation (SNSF) Grant No. CRSK-2$\_$196756. C.W.N. and C.M. acknowledge financial support by the Swiss National Science Foundation (SNSF) Grant No. P00P2$\_$170597. A.R. and H.H acknowledge financial support from the European Research Council (Grant ERC-2015-AdG-694097) and the Cluster of Excellence “CUI:Advanced Imaging of Matter” of the Deutsche Forschungsgemeinschaft (Grant EXC 2056 Project 390715994).
\end{acknowledgments}

\section{Experimental methods}

\begin{figure}[ht]
\includegraphics[width=0.75\columnwidth]{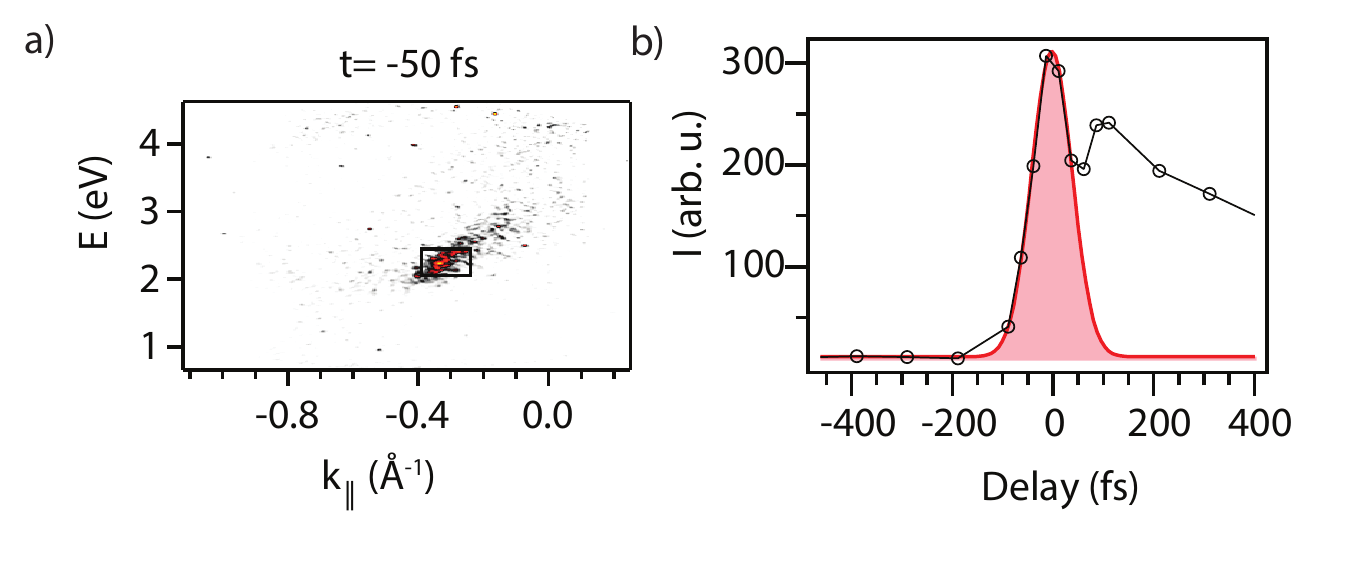}
\caption{\label{fig:SM1} a) ARPES intensity as a function of energy and parallel momentum showing the conduction states along the $\overline{\Gamma}-\overline{\Sigma}$ direction at a time delay of -50 fs. The pump-probe temporal cross correlation is determined by integrating the signal in the rectangular box. b) Temporal trace showing the integrated intensity in the box of panel a) as function of time. Red curve, Gaussian fit to the rising edge, the FWHM is 95 fs.}
\end{figure}

Commercial \wse\ single crystals where prepared by exfoliation in-situ under UHV conditions. The base pressure during the experiments was below 1$\times10^{-10}$ mbar. All the experiments were performed at room temperature, where no surface photovoltage or charging effects were observed.The light source is based on a high-harmonic generation of a high-repetition Ytterbium-based Optical parametric chirped pulse amplifier (OPCPA) \cite{Puppin2015}. The experiments were performed in an ARPES chamber equipped with a 6-axis manipulator and a hemispherical electron energy analyzer (Specs Phoibos 150), further details on the experimental setup are described in reference \cite{puppin_time-_2019}. The temporal time zero and pump probe cross correlation of 95 fs were measured by fitting the rising edge of the first observable signal in the excited-state band structure, as illustrated in Fig. \ref{fig:SM1}. The second maximum observed after 100 fs is a result of electron population scattered from other states during the energy relaxation process. 

\section{Data analysis}
In Fig. 1 e) of the main text, the experimental EDCs have been normalized to the same area as a function of parallel momentum in the VB. This was chosed for reducing the impact of matrix element in the display of constant energy map and for a clearer comparison with the \GW\ data. The same procedure was applied to EDCs in the CB (i.e. on the data for E$>$1 eV), but prior to the area normalization, an exponential background tail from the underlying occupied states was subtracted. 
No normalization procedure was performed on the data in Fig. 2, Fig. 3 and Fig. 4.

\section{Theoretical methods}

The electronic band structure of bulk \wse\ was computed using many-body perturbation theory at the one-shot \GW\ level on top of DFT results. This approach has been vastly employed in the literature for the description of the electronic properties of semiconductor materials due to its accuracy and good agreement with experimental measurements. The system was modelled using a hexagonal supercell with the experimenal lattice constants a = b = 3.28 \AA\ and c = 12.98 \AA \cite{el-mahalawy_thermal_1976}. DFT calculations were performed using the generalised gradient approximation (GGA) with the PBE functional\cite{perdew_generalized_1996}. The Brillouin zone (BZ) was sampled with a 9x9x9 k-point grid. We used a total of 1000 conduction bands and a 18 Ry energy cutoff for the computation of the inverse dielectric matrix. For the evaluation of the screened and bare Coulomb parts of the self-energy operator, we used energy cutoffs of 18 Ry and 160 Ry, respectively. Spin-orbit coupling was included directly in the DFT calculations and perturbatively at the \GW\ level, using the BerkeleyGW package \cite{deslippe_berkeleygw_2012}.  All employed cutoff values, BZ sampling and number of bands were systematically and independently increased until results were converged within few tens of meV for the conduction and valence band energy difference. Finally, we performed DFT calculations using a 24x24x9 BZ sampling and interpolated linearly the 9x9x9 GW band structure into this finer k-point grid.

\bibliography{biblio_EMAP}
\end{document}